# Odd-integer quantum Hall states and giant spin susceptibility in p-type few-layer WSe$_2$

*Shuigang Xu$^{1†}$, Junying Shen$^{1†}$, Gen Long$^1$, Zefei Wu$^1$, Zhi-qiang Bao$^2$, Cheng-Cheng Liu$^2$, Xiao Xiao$^1$, Tianyi Han$^1$, Jiangxiazi Lin$^1$, Yingying Wu$^1$, Huanhuan Lu$^1$, Jianqiang Hou$^1$, Liheng An$^1$, Yuanwei Wang$^1$, Yuan Cai$^1$, K. M. Ho$^1$, Yuheng He$^1$, Rolf Lortz$^1$, Fan Zhang$^{2*}$, Ning Wang$^{1*}$*

$^1$Department of Physics and Center for Quantum Materials, the Hong Kong University of Science and Technology, Clear Water Bay, Hong Kong, China

$^2$Departement of Physics, the University of Texas at Dallas, Richardson, Texas 75080, USA

$^†$These authors contributed equally

$^*$Corresponding author: zhang@utdallas.edu (F.Z.), phwang@ust.hk (N. W.)

## Abstract

**We fabricate high-mobility p-type few-layer WSe$_2$ field-effect transistors and surprisingly observe a series of quantum Hall (QH) states following an unconventional sequence predominated by odd-integer states under a moderate strength magnetic field. By tilting the magnetic field, we discover Landau level (LL) crossing effects at ultra-low coincident angles, revealing that the Zeeman energy is about three times as large as the cyclotron energy near the valence band top at Γ valley. This result implies the significant roles played by the exchange interactions in p-type few-layer WSe$_2$, in which itinerant or QH ferromagnetism likely occurs. Evidently, the Γ valley of few-layer WSe$_2$ offers a unique platform with unusually heavy hole-carriers and a substantially enhanced g-factor for exploring strongly correlated phenomena.**

Subjected to a sufficiently high magnetic field $B$, the Hall resistance of a two-dimensional electron gas (2DEG) undergoes QH transitions to take on the quantized values $h/\nu e^2$, where $h$ is the Planck's constant, $e$ is the elementary charge, and $\nu$ is the LL filling factor (FF). The effective Landé g-factor g$^*$ and the effective mass of carriers $m^*$ are two fundamental parameters that characterize the energy gaps of LLs. In the single-particle picture, the cyclotron energy $E_c = \hbar\omega_c = \hbar eB/m^*$, reflecting the quantization of an electron's orbital motion. The Zeeman energy $E_z = g^*\mu_B B$, where $\mu_B$ is the Bohr magneton, causes the spin splitting of the LLs. In GaAs based quantum wells, $E_c$ is much larger than $E_z$, and QH effects only occur at even-integer FFs in a low-field regime. At high fields, the electron-electron interactions [1-3] can substantially lift the spin degeneracy and give rise to odd-integer FFs. In 2DEGs with comparable $E_c$ and $E_z$, exotic QH states may appear. For instance, a series of even-denominator fractional QH states has been observed at a MgZnO/ZnO heterointerface [4,5] where $E_z/E_c = 0.95$. Here we report the surprising discovery of $E_z/E_c = 2.83$, the largest among all accessible 2DEGs to date, in p-type few-layer WSe2. This feature presents a remarkable 2DEG platform with strongly correlated phenomena.

The emergence of atomically thin crystals, such as graphene [6,7], black phosphorus [8,9], and transition metal dichalcogenides (TMDCs) [10,11], has greatly enriched the prospective platforms of 2DEGs. Among these materials, TMDCs, with strong spin-orbit coupling (SOC), large band gaps, and rich valley degrees [12], exhibit several extraordinary phenomena, such as circular dichroism [13-15], valley Zeeman effect [16-19], and opto-valley Hall effect [20]. However, because of contact problems and low carrier mobility, the investigation on TMDC QH transport has been limited until recently. In parallel with several significant efforts [10,11,21,22], two experiments observed the onset of integer QH states in TMDC devices: one for the *Q-valley electrons* [10] and the other for the *K-valley holes* [11]. In both cases, the transport features were predominated by *even-integer* QH states.

The band structures of TMDCs are thickness dependent. For WSe2, according to the first-principles calculations [23] and APRES experiments [24], the valence band maxima change from the K/K' points in monolayers/bilayers to the Γ point in thicker layers. Here, we report the surprising observation of an anomalous magnetotransport of *Γ-valley holes*, predominated by *odd-integer* QH states, in high mobility p-type few-layer WSe2. By means of tilt-field and temperature-dependent measurements, we further evidence LL crossing at ultra-low coincident angles and discover giant spin susceptibility $\chi^* = g^*m^*\mu_B/2\pi\hbar^2 \propto g^*m^* = 5.65m_0$, where $m_0$ is the electron mass. This remarkable characteristic is responsible for the observed unconventional FF sequence and amounts to an enormous ratio of Zeeman-to-cyclotron energies $E_z/E_c = g^*m^*/2m_0 = 2.83$. The heavy hole-carriers and the giant spin susceptibility suggest that p-type few-layer WSe2 is a fertile ground for exploring 2D strongly correlated phenomena.

To fabricate high-mobility p-type WSe2 devices, as illustrated in Figs. 1(a) and 1(b), we employ h-BN encapsulated structures and a dry transfer technique [22,25]. To access the Γ valley (p-type) of WSe2, we deposit a contact metal Pd through the selective etching technique [22,26]. The reason is that the work function of Pd (~5.6 eV) [27] matches the valence band edge of WSe2 (~5.2 eV) [28] and forms a low barrier with p-type WSe$_2$, which is confirmed by the $I_{ds}$-$V_{ds}$ curves (Fig. S1). The typical field-effect mobility $\mu_F$ achieved in our samples is about 220 cm$^2$/V s at room temperature and 12,000 cm$^2$/V s at 2 K. The Hall mobility $\mu_H$ ($\mu_H = \sigma/ne$ with $\sigma$ the sheet conductance and $n$ the carrier density determined by Hall measurements) is about 4,800 cm$^2$/V s ($n$ =4.1 × 10$^{12}$ cm$^{-2}$) at 2 K. Similar to other 2D semiconductors [29], the measured $\mu_H$ is lower than $\mu_F$, originating from the carrier density dependence of $\mu_H$.

Examples of longitudinal resistance ($R_{xx}$) and Hall resistance ($R_{xy}$) measured as functions of the magnetic field ($B$) at fixed carrier densities are shown in Fig. 1(c). At low $B$, pronounced Shubnikov-de Haas (SdH) oscillations are clearly visible. From the onset of SdH oscillations (3.5 T), we can estimate the quantum mobility $\mu_q \approx 1/B_q = 2,800$ cm$^2$/V s [30], further confirming the high quality of our sample. We can also extract the hole density participating in the oscillations using $n = g_s g_v e B_F/h$, where $g_s$ and $g_v$ are the spin and valley degeneracies, respectively, and $B_F$ is the oscillation frequency. $B_F$ can be extracted from fast Fourier transform (FFT) in Fig. 1(e). The carrier densities of our samples are rather low, and the Fermi energies are located near the valence band maxima, i.e., the Brillouin zone center (Γ point) with $g_s = 2$ and $g_v = 1$, as shown in Fig. 1(f). The hole densities obtained from the oscillations are found to be consistent with those obtained from Hall measurements (Fig. S3).

Similar to the characteristics of QH effects observed in other 2DEGs, as shown in Fig. 1(c), the appearance of quantized $R_{xy}$ plateaus is accompanied by the exhibition of local $R_{xx}$ minima with $R_{xx} \ll R_{xy}$ [31]. Counterintuitively, the QH states with odd-integer FFs, such as $\nu$ =19, 17, 15, and 13, already appear sequentially at low $B$, whereas the states with even-integer FFs, such as $\nu$ =14, 12, and 10, only appears at high $B$. A similar phenomenon was reported for the QH states in a polar oxide heterostructure [4,31]. The sequence of the observed FFs usually reflects $E_z/E_c$. As $E_z/E_c = m^*g^*/2m_0$, it provides a measure of the spin susceptibility $\chi^* \propto g^*m^*$. In GaAs or graphene 2DEGs [32,33], $E_z/E_c$ is normally in the order of 10$^{-2}$ because of the small $m^*$ and $g$-factor. Accordingly, the LLs with the same orbital index but opposite spins are nearly degenerate at low $B$, and only become split at sufficiently high $B$, when Zeeman effect is enhanced by electron-electron interactions [1-3]. Therefore, in these conventional systems, the even-integer QH states appear first, followed by odd-integer states, as the $B$ increases.

In light of the above analysis, the unconventional sequence of QH states in few-layer WSe$_2$ is strongly suggestive of $\chi^*$ being rather large. A simple analysis reveals that the odd-integer QH states predominate under low fields when $2j + 0.5 < E_z/E_c < 2j + 1.5$, with $j$ being a non-negative integer. Hence, we tilt $B$ to study the LL crossing, which has been a standard method to determine $E_z/E_c$ [2].

Under a tilted $B$, $E_c$ scales with the perpendicular field $B_\perp$, whereas $E_z$ scales with the total field $B_{tot}$. Therefore, tuning the tilt angle $\theta$ can increase the ratio $E_z/E_c$ and produce the crossing of LLs. As illustrated in Fig. 2(e), the evolution of the LL crossing is characterized by $i = E_z/E_c = g^*\mu_B B_{tot}/(\hbar e B_\perp/m^*)$. As $B_\perp = B_{tot}\cos\theta$ and $\mu_B = \hbar e/(2m_0)$,

$$i\cos\theta = g^*m^*/2m_0. \tag{1}$$

As $\theta$ increases, two LLs with opposite spins and orbital indices differing by $i$ cross each other at the so-called coincidence angles, where $i$ takes integer values.

Figure 2(a) displays the experiment data of the SdH oscillations at different tilt angles. For a fixed $\theta$, $R_{xx}$ and $R_{xy}$ are recorded as functions of $B_{tot}$. The value of $\theta$ is calibrated by the simultaneous $R_{xy}$ measurements in the non-quantized regime as the total carrier density does not change upon $\theta$. At $\theta = 0°$, the QH states with odd- and even-integer FFs are both observed at high $B$. As $\theta$ increases, the even-integer QH states at $\nu = $ 16, 18, and 20 at high $B$ become weaker and weaker and eventually vanish at the coincidence angle of 23.6°. Conversely, the odd-integer QH states change little. In other words, pronounced SdH minima (maxima) for the odd (even)-integer states are observed at $\theta = 23.6°$. When $\theta$ is tilted beyond 23.6°, the even-integer QH states gradually appear again at $\nu = $ 26, 28, and 30, whereas the odd-integer QH states at $\nu = $ 25, 27, and 29 gradually disappear. At the second coincidence angle of 45.9°, pronounced SdH minima (maxima) for the even (odd)-integer states are clearly observed, opposite to the case for $\theta$ = 23.6°. Likewise, the next-order phase reversal [34] in the SdH oscillations is observed at the coincidence angle of 54.8°.

We employ Eq. (1) to fit the three identified coincidence angles with respect to three adjacent integers in Fig. 2(b). Surprisingly, the coincidence angles for $\theta = $ 23.6°, 45.9°, and 54.8° correspond to $i = $ 3, 4, and 5, respectively. For GaAs, graphene, ZnO, and black phosphorus [4,8,35], the fitted $i$ values all start from 1. Here, if we assign the coincidence angle $\theta = 23.6°$ to $i = 1$ (or $i = 2$), the next-order coincidence angle will be $\theta = 62.7°$ (or $\theta = 52.3°$), which largely deviates from our observation $\theta = 45.9°$ in Fig. 2(a). Therefore, the fitting appears to be unique in light of Eq. (1). Markedly, with the identified $\theta$ and $i$ values, Eq. (1) yields $g^*m^* = 5.65m_0$, which is larger than those of the aforementioned 2DEGs. This result implies giant spin susceptibility in the p-type few-layer WSe$_2$.

Alternatively, we can also extract the value of $g^*m^*$ by analyzing the dependence of the SdH oscillation amplitudes on the tilt angles, as shown in Fig. 2(c). The amplitude of SdH oscillations can be described by the Lifshitz-Kosevich formula [29]:

$$\Delta R_{xx} = A_0 R_T R_D R_S \cos[2\pi \frac{B_F}{B_\perp} - \pi + \varphi] \text{ with } R_S = \cos(\frac{\pi}{2} \frac{g^* m^*}{m_0} \frac{B_{tot}}{B_\perp}), \quad (2)$$

where $A_0$ is a constant, $R_T$ is the temperature factor, $R_D$ is the Dingle factor, and $R_S$ is the spin damping factor of Zeeman splitting. The Berry phase $\varphi$ is found to be zero (Fig. S4), which is consistent with the single-band picture revealed below by our first-principles calculations. At fixed temperature and $B_\perp$, the $B_{tot}$ dependence of the SdH amplitude is determined [36] by $R_S$. As shown in Fig. 2(d), the observed SdH amplitudes as functions of $1/\cos\theta = B_{tot}/B_\perp$ at $v = 25$ and 26 are well reproduced by Eq. (2) with $g^*m^* = 5.93m_0$ for v=25 and $g^*m^* = 5.84m_0$ for v=26.

The effective hole mass $m^*$ can be determined by the temperature dependence of SdH oscillations. The temperature-induced damping in Eq. (2) is [26,29] $R_T = \xi/\sinh(\xi)$, where $\xi = 2\pi^2 k_B T/\hbar\omega_c$ and $k_B$ is the Boltzmann constant. The SdH oscillations at different temperatures, for a fixed carrier concentration and $B < 9$ T, are shown in Fig. 3(a). The corresponding FFT confirms the single SdH oscillation frequency. By fitting $\Delta R_{xx}$ as a function of $T$ using $R_T$, we obtain the hole cyclotron mass $m^* = 0.89m_0$, indicating $g^* = 6.35$.

The band structures of few-layer WSe2 are thickness dependent. Different from those at K valleys, the $m^*$ at $\Gamma$ valleys of WSe2 has been confirmed to vary from $0.5m_0$ in 3D bulk to $2.8m_0$ in monolayers [24,37-39]. Since our samples are about 8-layer thick, we perform first-principles calculations to examine the band structure of 8-layer WSe2, as shown in Fig. 1(f). Our calculated $m^*$ is isotropic and ~ $0.76m_0$, which is consistent with our measurement. As the relevant sub-band mainly arises from the delocalized out-of-plane $5d_z^2$ orbitals of W atoms, the SOC is negligibly weak, thus implying $g^* = 2$ according to Roth's formula [40]. Apparently, this band structure result is smaller than the experimental value of $g^*$. The discrepancy is likely due to the exchange interactions of holes, as will be elaborated in below.

Having established that at a zero tilt angle $E_z/E_c = g^*m^*/2m_0 = 2.83$, we conclude that, surprisingly, the LL crossing among those with different orbital indices already exists when $\theta = 0°$. Specifically, as $2.5 < E_z/E_c < 3.5$, the integer QH plateaus should follow the sequence of $v = 1, 2, 3, 5, 7, 9, 11 \ldots (2M+1) \ldots$ at small fields, in which the *N*-th LL with a downward spin and the (*N+3*)-th LL with an upward spin are nearly degenerate (Fig. 4(a)). The gap between the two nearly degenerate LLs scales linearly with *B* and becomes sufficiently large at high fields, in which the even-integer QH states with large orbital indices can be fully resolved. The QH plateaus exhibited in Fig. 1 are consistent with such a physical picture.

According to the LL structure in Fig. 4(a), the energy gaps at odd ($v$ = 3, 5, 7, …) and even ($v$ = 4, 6, 8, …) FFs can be respectively expressed by

$$\Delta_1 = E_z - 2E_c - \Gamma \quad \text{and} \quad \Delta_2 = 3E_c - E_z - \Gamma, \tag{3}$$

where $\Gamma$ is the LL broadening caused by disorder. To determine the energy gaps, we measure the temperature dependence of $R_{xx}$ as a function of $V_g$ at a fixed $B$. Figure 4(b) displays the typical results for the $v$ = 11 and 12 states at $B$ = 13 T. Figure 4(c) and 4(d) further plot the $1/T$ dependence of $R_{xx}$ minima at various $B$. As the energy gaps $\Delta_1$ and $\Delta_2$ can be approximated to the corresponding activation gaps, we deduce them by considering $R_{xx} \propto \exp(-\Delta/2k_BT)$, where the activation gap $\Delta$ can be extracted from the Arrhenius plots in Fig. 4(c) and 4(d). As clearly shown in Fig. 4(e), the measured activation gaps at $v$ = 11 and 12 both scale linearly with the $B$. By relating the two fitted lines in Fig. 4(e) to Eq. (3), we obtain $E_c$ = 1.69$B$ K and $E_z$ = 4.38$B$ K, which lead to an effective mass $m^* = 0.80m_0$ and a g-factor $g^* = 6.52$. These results are consistent with those obtained from the coincident technique and Dingle plot.

The previously reported TMDC QH states in n-type few-layers and p-type mono- and bi-layers are in sharp contrast to the ones studied here in p-type few-layers. In the former case, the QH transport emerges from the six $Q$ valleys [10,22], where strong SOC and inversion symmetry (asymmetry) lead to $g_s$ = 2 (1) for even (odd) layers. In the latter case, the QH transport was attributed to the two $K$ valleys [11,41]. In the present case, however, the single $\Gamma$ valley with $g_s$= 2 and negligible SOC is the one that governs the QH transport. The observed QH states were predominantly at even-integer FFs in the previous two, whereas the odd-integer QH states dominate in the present study. This unexpected observation uncovers the giant spin susceptibility $\chi^* \propto g^*m^* = 5.65m_0$ in p-type few-layer WSe$_2$.

Given the weak interlayer couplings in WSe$_2$, one may wonder whether or not only the top-layer carriers participate in the quantum transport in our devices. The odd-integer QH states are predicted to predominantly occur in p-type monolayers [41], in which the mirror symmetry and the strong SOC dictate a large out-of-plane spin splitting. However, this explanation can be excluded, as the SdH oscillations in Fig. 2(a) are highly sensitive to $B_{tot}$ at fixed $B_\perp$. This analysis leaves the electron-electron interactions as the seemingly only possible explanation for the enhanced g-factor. For long-range interactions, the exchange energy decreases rapidly with increasing LL orbital index and scales with $B_\perp^{1/2}$, different with the orbital independence and the linear scaling in Fig. 2b. The 5$d$ orbital relevant for $\Gamma$-valley holes is in fact more localized than with the $p$ orbitals in GaAs or graphene. It is thus reasonable to speculate on the short-range interactions, the exchange energy of which scales with $B_\perp$ for each occupied LL. Nevertheless, the observed odd-integer states are all QH ferromagnets with uneven populations of spin-up and -down LLs.

Two implications follow from the determined large effective mass $m^*$ for $\Gamma$-valley holes of WSe$_2$. First, the cyclotron energy is reduced by a factor of approximately 10, compared with the typical value in GaAs-based 2DEGs. This implication explains why higher mobilities are required to observe the onset of QH effects in TMDC 2DEGs. Second, the density of states, which is linearly proportional to $m^*$, and the interactions of 5$d$ holes are both substantial. Following the Stoner criterion, itinerant ferromagnetism may exist in p-type few-layer WSe$_2$, thus suggesting an already established Zeeman splitting at $B = 0$. This possibility is indeed consistent with our observed coincidence angles, and yields the same value of $g^*m^*$, if the integer $i$'s in Eq. (1) are fitted to 1, 2, and 3 or 2, 3, 4, instead of 3, 4, and 5 (Fig. S7).

In summary, we observed an unconventional sequence of $\Gamma$-valley QH states in high mobility WSe$_2$, predominantly at odd-integer FFs. The observed Hall plateaus, LL crossing, SdH amplitudes, and activation gaps are in good harmony with each other, discovering the giant spin susceptibility $\chi^* \propto g^*m^* = 5.65m_0$. This feature implies the significant roles played by the electron-electron exchange interactions in p-type few-layer WSe$_2$, in which itinerant ferromagnetism or QH ferromagnetism likely occurs. Remarkably, the $\Gamma$ valley of few-layer WSe$_2$ offers an unprecedented 2DEG platform with unusually heavy hole-carriers and a large g-factor for exploring strongly correlated phenomena.

**Acknowledgement**

We acknowledge the financial support from the Research Grants Council of Hong Kong (Project Nos. 16302215, HKU9/CRF/13G, 604112 and N_HKUST613/12) and the UT-Dallas Research Enhancement Funds. We are grateful for the technical support of the Raith-HKUST Nanotechnology Laboratory for the electron-beam lithography facility at MCPF. F.Z. is grateful to Xiaoyan Shi for the valuable suggestions on the tilt field measurement and to Kavli Institute for Theoretical Physics for hospitality during the finalization of this work, which was supported in part by NSF Grant No. PHY11-25915.

**Figure captions**

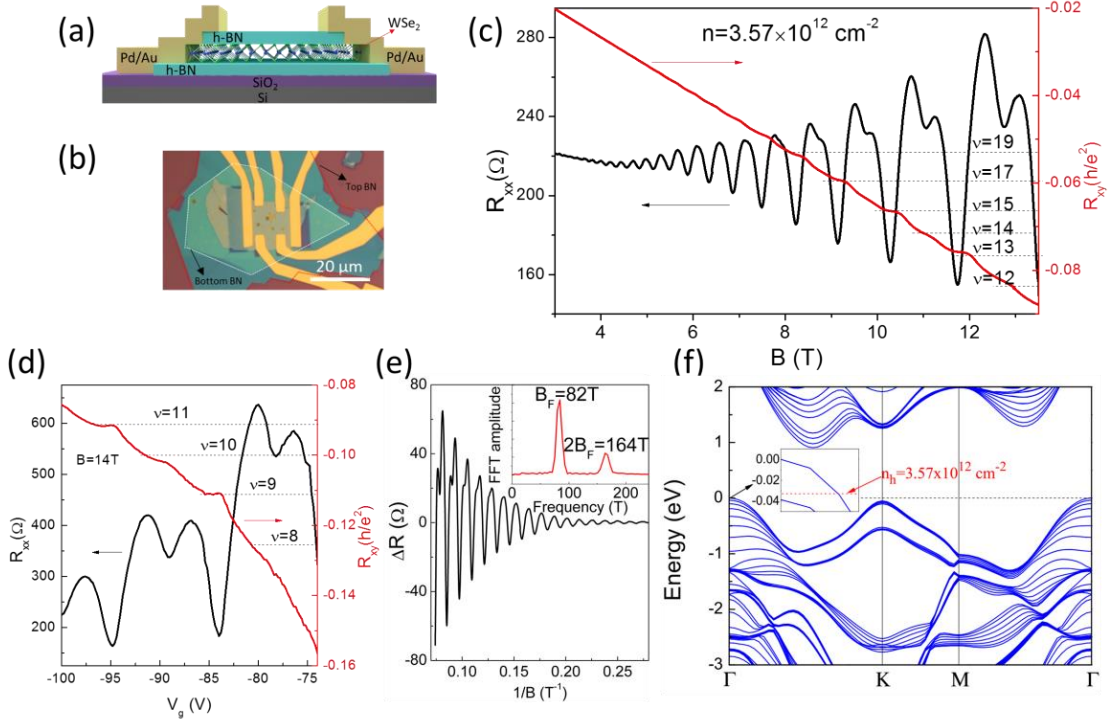

FIG. 1. (a) Schematic of h-BN encapsulated WSe$_2$ structures. (b) Optical image of a Hall device in our measurement. (c) $R_{xy}$ and corresponding $R_{xx}$ as functions of $B$ measured at 1.7 K for a hole density of 3.57×10$^{12}$ cm$^{-2}$. (d) $R_{xy}$ and $R_{xx}$ as functions of $V_g$ at 14 T. The dashed lines mark the values of the quantized Hall plateaus ($h/\nu e^2$). (e) Background subtracted magnetoresistance as a function of $1/B$ at n=3.57×10$^{12}$ cm$^{-2}$. The inset is the corresponding FFT. (f) Calculated band structure of 8-layer WSe$_2$. The Fermi level for (c) crosses only the highest spin-degenerate sub-band at the $\Gamma$ valley.

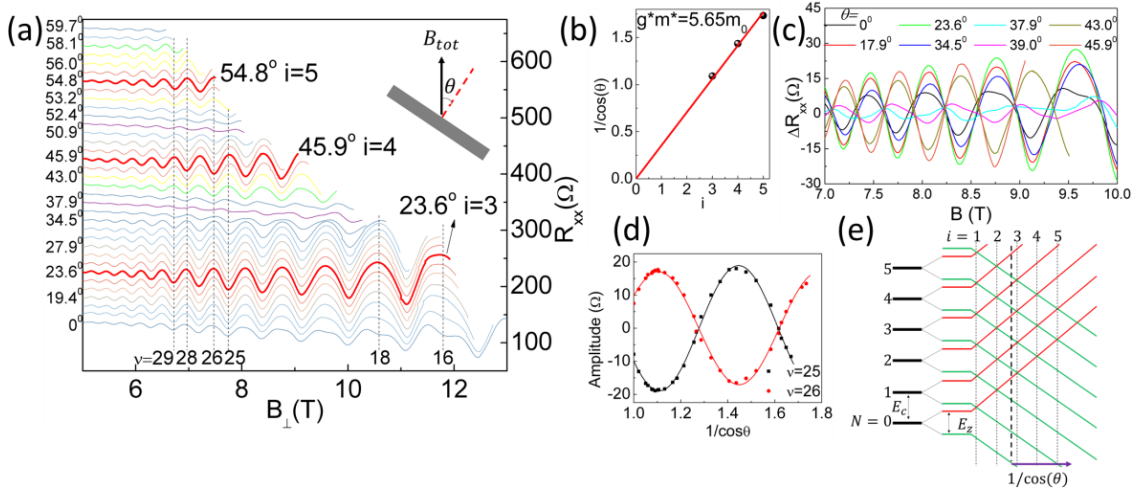

FIG. 2. (a) $R_{xx}$ as a function of $B_\perp$ at various tilted angles at 1.7 K. The inset shows the illustration of the angles. Data were recorded at a fixed $V_g$ and shifted vertically for clarity. Coincidences at $i = 3, 4, 5$ are emphasized as bold lines. (b) $1/cos\theta$ of the identified coincidence angles as a function of $i$. (c) Comparison of SdH amplitudes at various tilted angles from the first ($i = 3$) to the second ($i = 4$) coincident angle. Smooth backgrounds were subtracted for each line. (d) Evolution of SdH oscillation amplitudes as a function of $1/cos\theta$ at FFs of $\nu = 25$ and 26. The solid curves are the fitted cosine functions in Eq. (2). (e) Schematic of spin-split LLs as a function of the tilted angle for a constant $B_\perp$. The black dash line marks the location of $\theta = 0°$.

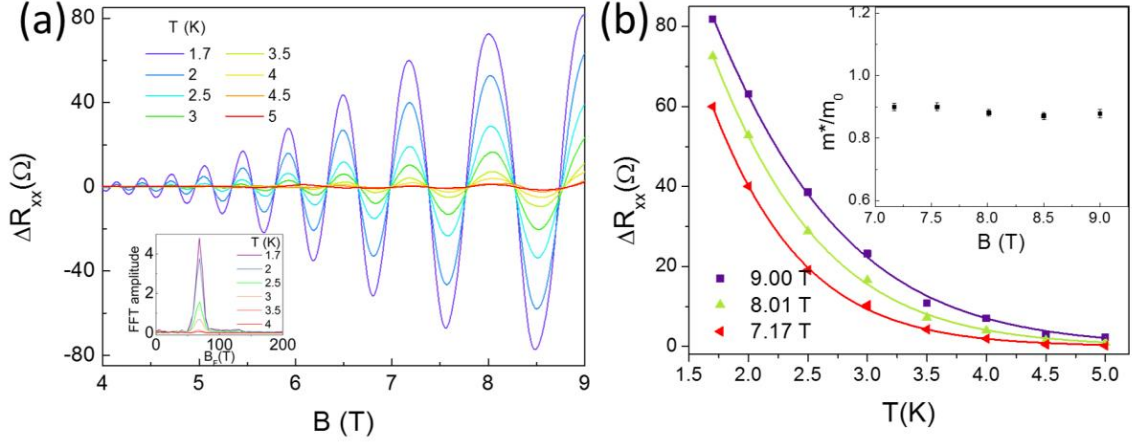

FIG. 3. (a) $\Delta R_{xx}$ as a function of $B$ measured at various temperatures with fixed $V_g$. The inset is the FFT corresponding to $\Delta R_{xx}$ vs $1/B$ data. The single SdH oscillation frequency confirms no observed Zeeman splitting at low $B$. (b) SdH oscillation amplitudes as functions of temperature at different $B$. The solid lines are the fitted curves of $R_T$. The inset shows the fitted hole effective masses at various $B$.

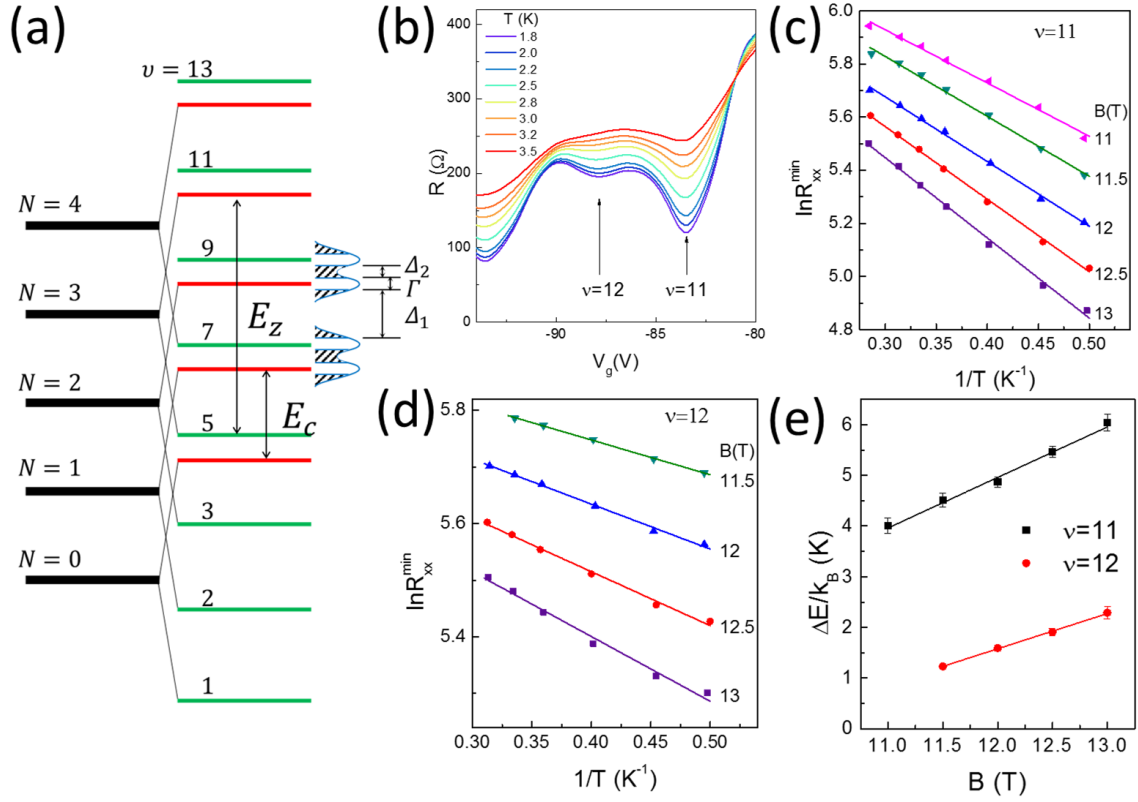

FIG. 4. (a) Schematic of LLs at the zero tilt angle. The activation gaps at the odd and even FFs ($\Delta_1$ and $\Delta_2$, respectively) with the LL broadening $\Gamma$ are marked. (b) $R_{xx}$ as a function of the $V_g$ at $B$=13 T at different temperatures. The arrows mark the $\nu$ =11 and 12 minima. (c) and (d) Arrhenius plots of $R_{xx}$ for the gaps at $\nu$ =11 and 12 at different $B$. Linear fits of the data yield the corresponding gaps $\Delta_1$ and $\Delta_2$. (e) Energy gaps as functions of $B$ at $\nu$ =11 and 12 deduced from the Arrhenius plots. The linear fits yield the effective mass and g-factor.

# Supplementary Material

1. **Experimental methods**

Bulk crystals of WSe$_2$ were grown by chemical vapor transport using I$_2$ as the carrier gas. The h-BN encapsulated WSe$_2$ devices were fabricated by the dry transfer technique. In a typical process, a top h-BN exfoliated on PMMA was used to pick up a few-layer WSe$_2$ flake and then transfer it on another bottom h-BN. To fabricate the electrodes, the top h-BN was selectively etched by O$_2$ plasma via the reaction ion etching with a controlled etching rate. To achieve p-type WSe$_2$ devices, Pd/Au (20/80 nm) was used as the contact metals by means of the standard e-beam lithography and the e-beam evaporation technique. The magneto transport measurements were conducted using the ac lock-in amplifier techniques at a cryostat temperature. The sample was loaded on an *in situ* rotating stage to allow for the tilted field studies.

2. **Output characteristics of p-type WSe$_2$ devices contacted with Pd.**

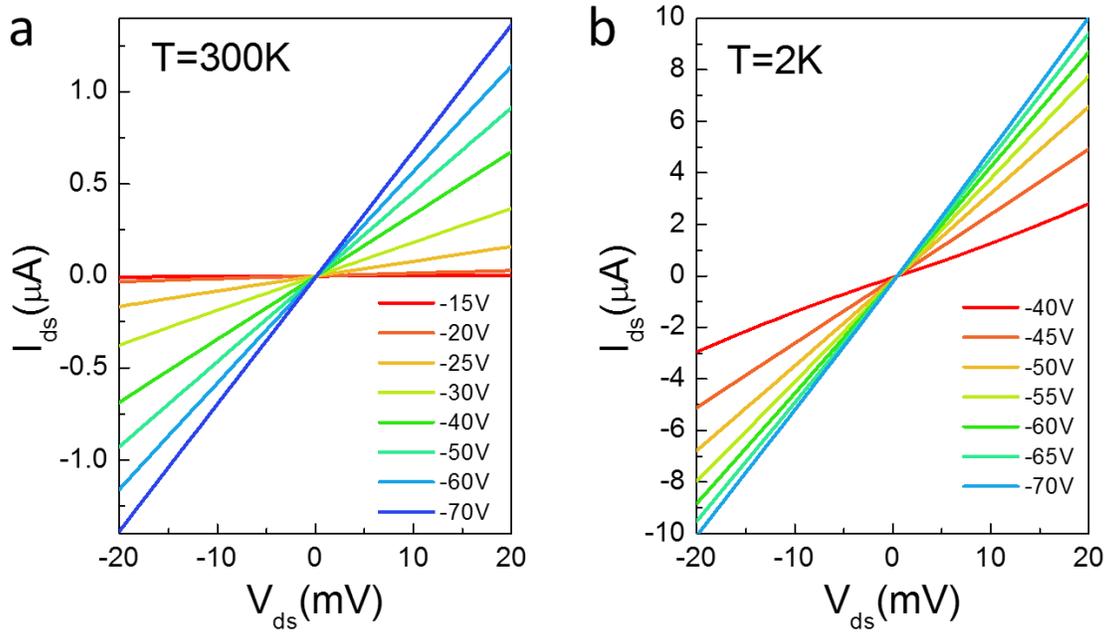

FIG. S1. I-V curves of p-type WSe$_2$ based field-effect transistors for various back gate voltages at temperature of (a) 300 K and (b) 2K. The linear I-V behaviors observed in these devices indicate that the contacts are Ohmic contact. The linear behavior can hold down to cryostatic temperature at high gate voltages.

### 3. Field-effect mobility

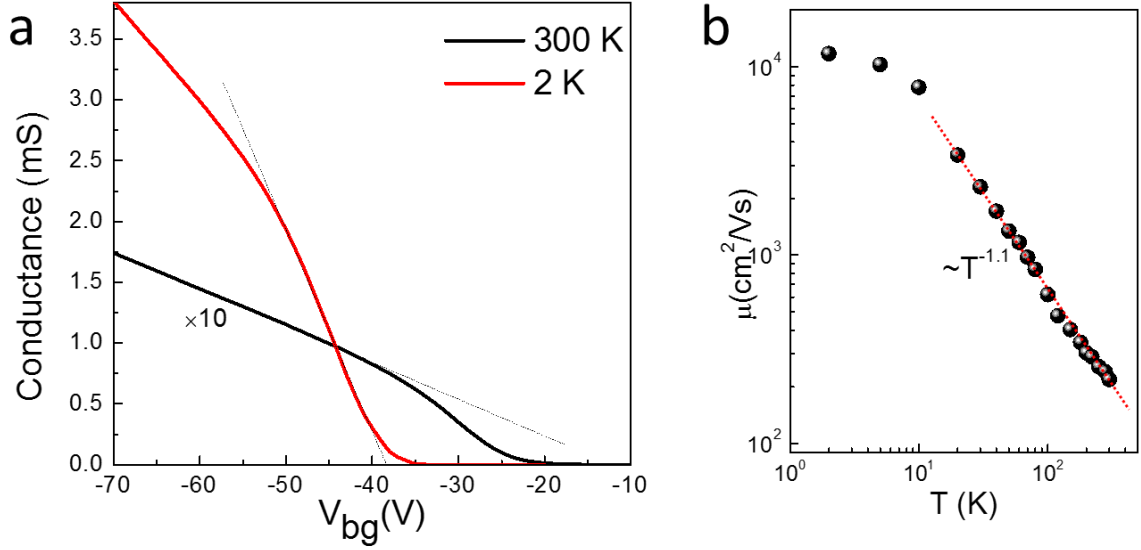

FIG. S2. Transfer characteristics of p-type WSe$_2$ devices contacted by high work function metal Pd. (a) Conductance as a function of back gate voltages at 300 K and 2 K. To clarify, the curve recorded at 300 K is enlarged by 10 times. The fit linear dash lines were used to extract the field-effect mobility by $\mu_F = (d\sigma/dV_g)/C_g$, where $\sigma$ is the sheet conductance. The capacitance of the 300 nm SiO$_2$ is $1.3\times10^{-8}$ F/cm$^2$, which is calibrated by Hall measurement. The calculated hole field-effect mobility in our best sample is up to 220 cm$^2$/V s at room temperature and dramatically increases to 12,000 cm$^2$/V s at 2 K. (b) The temperature dependent field-effect mobility. The linear fit in the phonon dominated regime is marked.

## 4. Carrier density

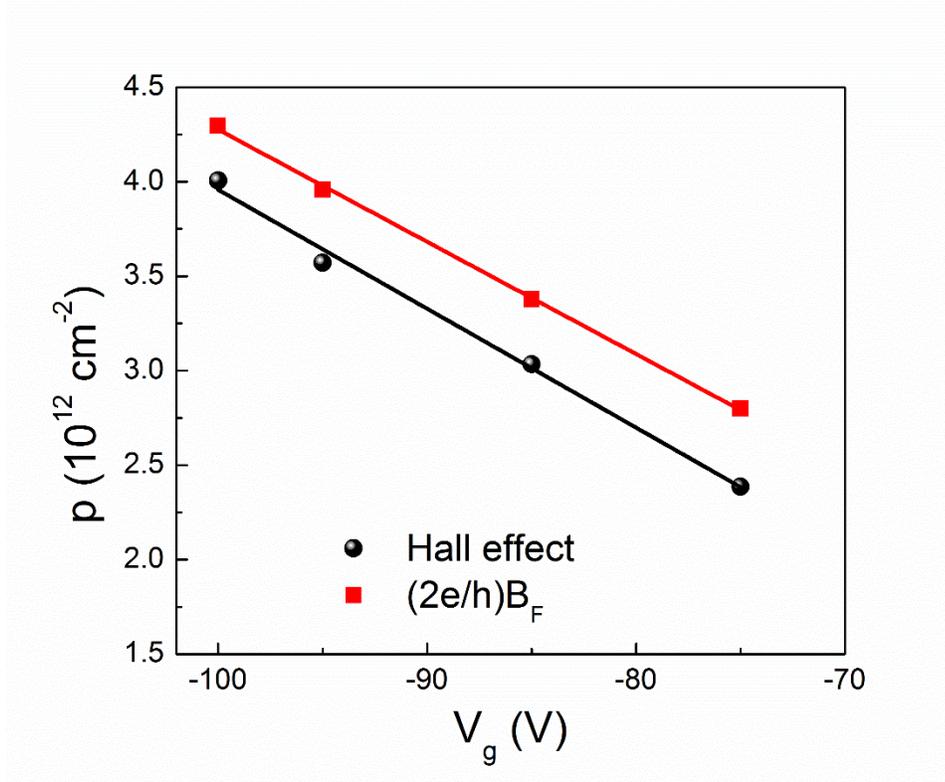

FIG. S3. Comparison of hole carrier densities measured by Hall effect and SdH oscillations at various gates. The black circles are calculated from the slopes of $R_{xy}$ dependent on magnetic field by Hall measurement. The red rectangles are calculated from the periods of SdH oscillations in $R_{xx}$. Twofold Landau level degeneracy with $g_s = 2$ and $g_v = 1$ is considered. The two kinds of densities are consistent with each other within the error range of 10% tolerance.

## 5. Berry phase

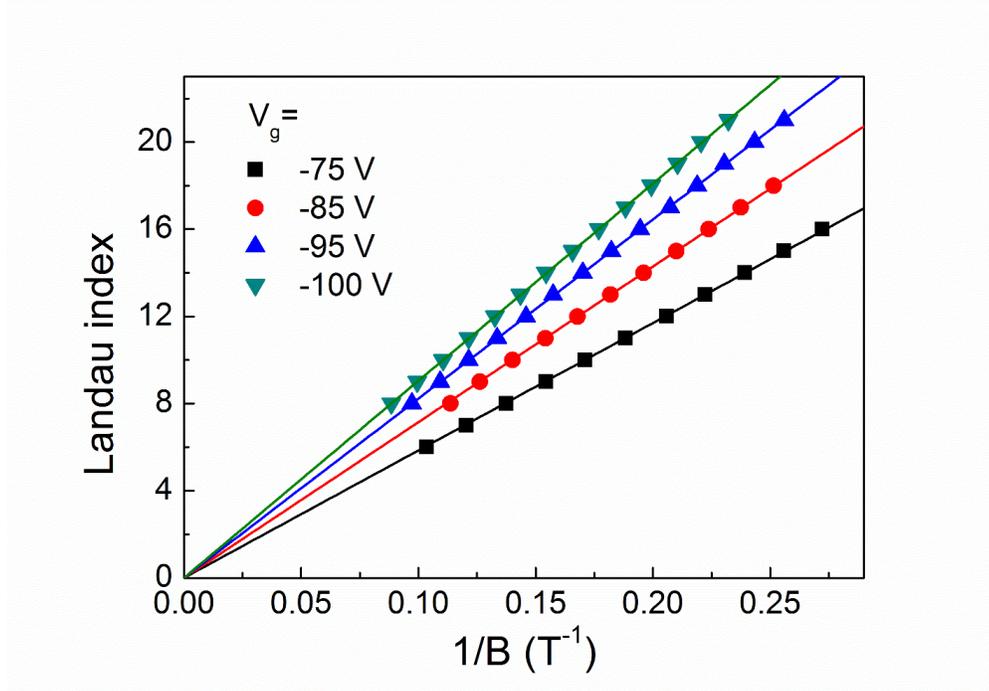

FIG. S4. Landau level fan diagram for SdH oscillations at various gate voltages. The locations of $1/B$ for the $n$th minimum of $R_{xx}$ are plotted against Landau level index $n$. The lines correspond to linear fits. The $n$-axis intercept provides a zero Berry phase in $\Gamma$ valley of valence band of WSe$_2$.

## 6. Calibration of Hall resistance

Due to the imperfect geometry in some devices, the measured Hall resistance $R_{xy}$ may contains some components from $R_{xx}$. To avoid this, $R_{xy}$ was recorded at both positive and negative magnetic fields. The calibrated $R_{xy}$ is obtained by $[R_{xy}(B) - R_{xy}(-B)]/2$.

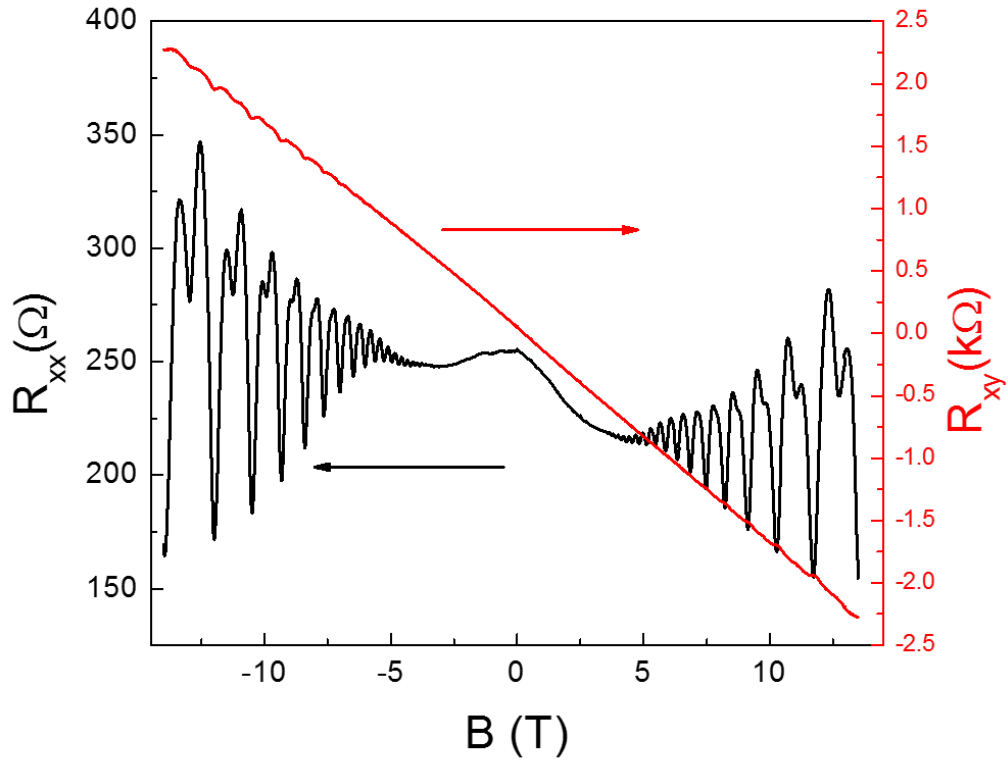

FIG. S5. Longitudinal and transverse resistance as a function of magnetic fields recorded at both negative and positive magnetic fields.

7. **Landau level crossing in fixed total magnetic field**

Additional to the measurements in Fig. 2, the coincident technique can also be achieved by measuring magnetoresistance as a function of back gate voltages at fixed total magnetic fields as shown in Fig. S6. In this case, when we tilt the sample stage, the Zeeman energy is unchanged since the total field is fixed. The cyclotron energy is decreased as $E_c = \hbar e B_{tot} cos\theta/m^*$. Similar as that in Fig. 2, we observed the coincident angle at 23.0°, 46.3° and 54.3°. The fitting results in Fig. S6b yield $g^* m^* = 5.68 m_0$, which agrees well with the coincident measure method in Fig. 2.

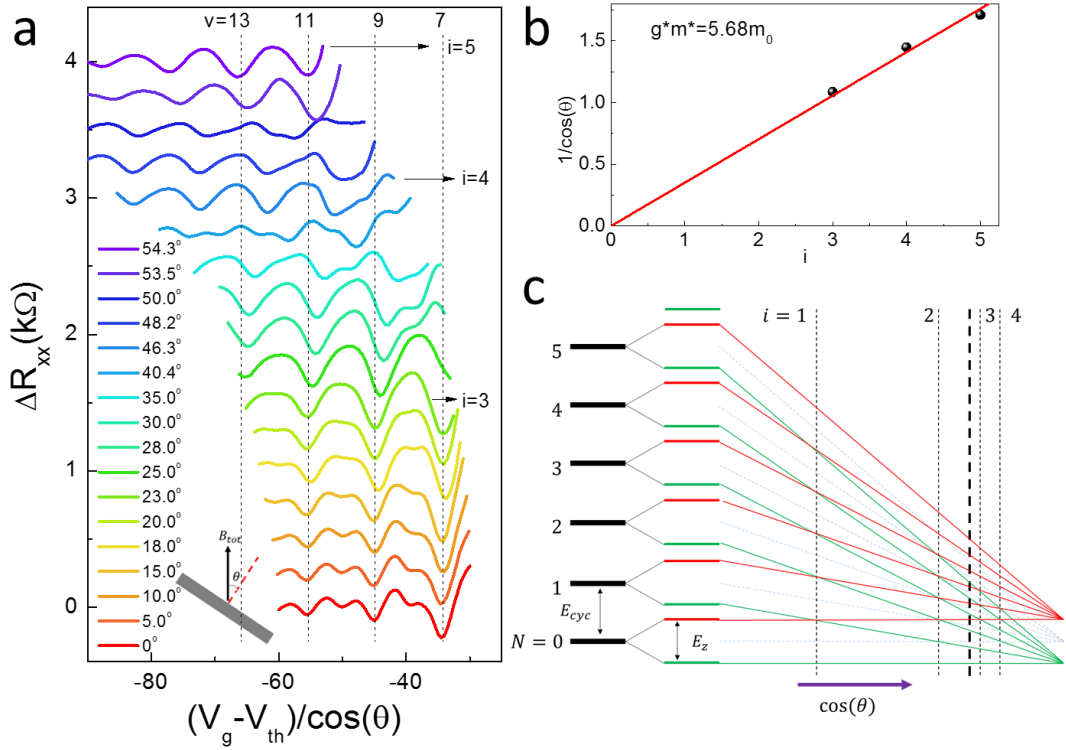

FIG. S6. Detecting Landau level crossing at fixed total magnetic fields by coincidence technique. (a) the magnetoresistance (after subtracting a smooth background) as a function of $(V_g - V_{th})/\cos\theta$ at various tilt angles. $V_g$ is the applied gate voltage. $V_{th}$ is the threshold voltage. The total magnetic field is fixed at 13.5 T. The data are recorded at 1.5 K. The tilt angle $\theta=0°$ is corresponding to the status of the field perpendicular to sample surface. (b) $1/\cos\theta$ of the identified coincidence angles as a function of $i$. The slope reflects the $g^*m^*$ of the system. **c**, Schematic diagram showing spin split Landau levels as a function of tilted angle. Since the total magnetic field is fixed, when tilting sample, the Zeeman splitting is constant, while the cyclotron energy gap is decreased. The black dash line marks the location of the studied system at $\theta=0°$.

## 8. Precise determination of tilted angles in coincident methods

The tilted angle can be calibrated by two methods[1]. The first calibration method is the Hall resistance measurement. The total carrier density $n_0$ does not change while tilting, while the measured carrier density by Hall effect ($n_\theta = 1/e(dR_{xy}/dB)$) at tilting angle of θ is $n_\theta = n_0\cos\theta$. The total carrier density can be measured at zero tilt angle by $n_0 = 1/e(dR_{xy}/dB)$). Then the calibrated angle is $\cos\theta = n_\theta/n_0$. The angles in Fig. 2 are calibrated by the first method.

The second is the period of the oscillation[2]. For 2DEG, each Landau level contains carrier density of $N = 2eB_\perp/h$. The gate induced carrier density is $n_s = C_g(V_g - V_{th})$, where $C_g$ is the capacitance and $V_{th}$ is the gate voltage threshold. The Landau level degeneracy is then $n_s = C_g\Delta V_g = 2eB\cos\theta/h$, where $\Delta V_g$ is oscillation period in gate voltage. Then the tilted angle can be calibrated by $\cos\theta = \Delta V_g/\Delta V_0$, where $\Delta V_0$ is the oscillation period at θ = 0°. The angle in Fig. S6 is calibrated by the second method.

## 9. Possibility of itinerant ferromagnetism

If at *B*=0, the electron-electron interactions already lead to itinerant ferromagnetism, then there is an already established Zeeman term at B=0. In this case, the coincidence relation is described by $i = (E_z + \Delta)/E_c = (g^*\mu_B B_\perp/\cos\theta + \Delta)/(\hbar eB_\perp/m^*)$, where Δ is a constant. We can fit the coincidence angles using linear equation $\frac{1}{\cos\theta} = \frac{2m_0}{g^*m^*}i - \Delta m^*/\hbar eB_\perp \frac{g^*m^*}{2m_0}$ by considering *i*=1, 2, 3 or 2, 3, 4 at θ=23.6°, 45.9° and 54.8°. The fits are shown in Fig. 7S. The slopes yield the same value of $g^*m^*$ as that discussed in main text.

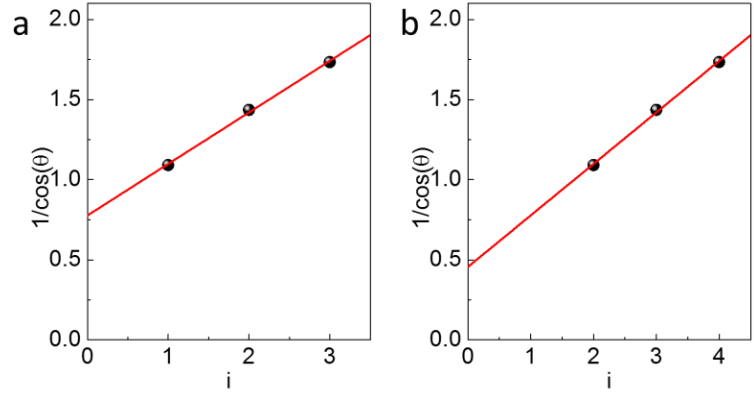

FIG. S7. Coincidence relations by considering itinerant ferromagnetism. $1/cos\theta$ of the identified coincidence angles as a function of $i$, for **a**, $i$=1, 2, 3 and **b**, $i$=2, 3, 4.